\documentclass[
reprint,
superscriptaddress,
amsmath,amssymb,
aps,
pra,
author-numerical,
nofootinbib,
]{revtex4-2}

\usepackage{graphicx}
\usepackage{dcolumn}
\usepackage{bm}
\usepackage{xcolor}
\usepackage{bbm}
\usepackage{physics}
\usepackage{appendix}
\usepackage{soul}
\usepackage{epsfig}
\usepackage{float}
\usepackage{multirow}
\usepackage{braket}
\usepackage{lipsum}
\usepackage{hyperref}
\hypersetup{
    colorlinks=true,
    linkcolor=blue,
    filecolor=magenta,      
    urlcolor=blue,
    citecolor = blue,
    }
\usepackage[normalem]{ulem}
\begin{document}
\preprint{APS/123-QED}

\title{Optimum parameter estimation of shaped phase objects} 

\author{Arturo Villegas}
\affiliation{ICFO—Institut de Ciencies Fotoniques, the Barcelona Institute of Science and Technology, 08860 Castelldefels (Barcelona), Spain}
\affiliation{Centre Tecnol\`ogic de Telecomunicacions de Catalunya (CTTC/CERCA), Av. Carl Friedrich Gauss, 7, 08860, Barcelona, Spain}
\email{arturo.villegas@alumni.icfo.eu}

\author{M. H. M. Passos}
\affiliation{ICFO—Institut de Ciencies Fotoniques, the Barcelona Institute of Science and Technology, 08860 Castelldefels (Barcelona), Spain}

\author{Silvania F. Pereira}
\affiliation{Department of Imaging Physics, Faculty of Applied Sciences, Delft University of Technology, Lorentzweg 1, 2628 CJ Delft, The Netherlands}

\author{Juan P. Torres}
\affiliation{ICFO—Institut de Ciencies Fotoniques, the Barcelona Institute of Science and Technology, 08860 Castelldefels (Barcelona), Spain}
\affiliation{Department of Signal Theory and Communications, Universitat Politecnica de Catalunya, 08034 Barcelona, Spain}
\email{juanp.torres@icfo.eu}

\date{\today}

\begin{abstract}
We show a general method to estimate with optimum precision, i.e., the best precision determined by the light-matter interaction process, a set of parameters that characterize a phase object. The method derives from ideas presented by Pezze et al., [Phys. Rev. Lett. {\bf 119}, 130504 (2017)]. Our goal is to illuminate the main characteristics of this method as well as its applications to the physics community, probably not familiar with the {\it formal} quantum language usually employed in works related to quantum estimation theory. First, we derive precision bounds for the estimation of the set of parameters characterizing the phase object. We compute the Cr\'amer-Rao lower bound for two experimentally relevant types of illumination: a multimode coherent state with mean photon number $N$, and $N$ copies of a multimode single-photon quantum state. We show under which conditions these two models are equivalent. Second, we show that the optimum precision can be achieved by projecting the light reflected/transmitted from the object onto a set of modes with engineered spatial shape. We describe how to construct these modes, and demonstrate explicitly that the precision of the estimation using these measurements is optimum. As example, we apply these results to the estimation of the height and sidewall angle of a cliff-like nanostructure, an object relevant in semiconductor industry for the evaluation of nanofabrication techniques.
\end{abstract}

\maketitle

\section{Introduction}
\label{Sec:Intro}
Phase objects are samples that add a spatially-dependent phase shift to the light that passes through them, or is reflected by them, while causing very low absorption. They are ubiquitous in many different areas of science and technology. Many biological specimens act as phase objects and techniques such as phase-contrast or differential interference contrast microscopy \cite{Murphy2012Book} have been developed to examine those specimens. In the semiconductor industry, the characterisation of printed phase objects such as steep steps, gratings and cliff-like structures are used in the evaluation of the performance of lithographic or other nanofabrication techniques \cite{raymond,roman,silvania2}.

In an imaging scenario, the aim is to retrieve a full image of the sample with the best spatial resolution available. This turns out to be challenging when one considers increasingly smaller structures, i.e. samples with sub-wavelength spatial features. For specific applications, several techniques have been found such as the use of shorter wavelengths for illumination (Extreme Ultra Violet and X-ray) \cite{xray1,xray2}, super-resolution techniques that make use of fluorophores \cite{hell1,hell2}, or by measuring in the near-field regime at distances from the sample where one can exploit the evanescent
fields \cite{nearfield1,nearfield2,nearfield3}. However, these techniques also show specific drawbacks such as the use of wavelengths, that allow low penetration depth into the samples and might cause damage to them, the need to use external markers that contaminate the samples or the requirement of unbearable exposure times due to scanning of large samples.

In many cases one is interested in estimating certain relevant parameters of a sample, for instance the period of a submicrometer grating \cite{silvania3} or the thickness of a thin nanolayer \cite{nathaniel1}. In this parameter estimation scenario, one wonders what is the minimum size of spatial features that can be estimated, i.e.,  what is the best spatial resolution achievable. It is also important to search for measurement schemes that can attain such a fundamental limit, which we will refer to as optimum measurement. Quantum estimation theory \cite{helstrom,holevo} provides the tools to answer these questions. Although the relevance of quantum estimation theory for determining the resolution limits in sensing and metrology has been known for some time \cite{helstrom1970estimation,fujiwara1994one,fujiwara1994multi,matsumoto,matsumoto2002}, it has only become of widespread use in optical sensing and imaging during the past few years \cite{tsang2015,tsangPRX,lupo,paris,yangfisher,vrehavcek2018optimal}.

We consider a scenario where the probe beam, after reflection/transmission from the phase object, can be written as a pure state
$|\Phi(\theta) \rangle$, where $\theta \equiv \left\{ \theta_1,\theta_2 \hdots \theta_M \right\}$ is the set of $M$ parameters that we want to estimate. A key tool in quantum estimation theory is the $M \times M$ quantum Fisher information matrix (QFIM), $F_{ij} = 4 \,\Omega_{ij}$, where
\begin{equation}
    \Omega_{ij} = \Re \left\{ \bra{\Phi_i} \Phi_j\rangle+\bra{\Phi_i}\Phi\rangle \bra{\Phi_j}\Phi \rangle \right\}, \label{fisher_pure}
\end{equation}
and $\ket{\Phi_i} \equiv \ket{\partial \Phi/\partial \theta_i}$. It allows to calculate the Quantum Cr\'amer-Rao bound (QCRB), a precision bound to the values of the elements of the covariance matrix  $V(\hat{\theta})$ for an unbiased estimator $\hat{\theta}$ of the set of parameters $\theta$, so that the matrix $[V(\hat{\theta})-F^{-1}]$ should  be positive semidefinite \cite{fujiwara1994multi}. This is a fundamental precision bound that is independent of the estimator and the specific measurement scheme considered, whose value depends only on the properties of the illumination light, beam, i.e., its quantum state, and the nature of the light-matter interaction that modifies the characteristics of the quantum state.  The QCRB is the most informative bound in the case of one-parameter estimation. This is also the case for multiparameter estimation if \cite{matsumoto2002,pezze2017}
\begin{equation}
\label{matsumoto}
\Im\, \bra{\Phi_i}\Phi_j\rangle=0 \Longleftrightarrow \bra{\Phi_i}\Phi_j\rangle=\bra{\Phi_j}\Phi_i\rangle.
\end{equation}
In some cases of interest the theoretical analysis of the QFIM is made considering that the source of light consists of $N$ copies of a multimode single-photon quantum state, even though the light source does not actually generate single-photon quantum states. For instance, in Ref. \cite{adesso2019} they justify using single-photon quantum states for analysing weak thermal sources at optical frequencies by claiming that the source is  \textit{"effectively emitting at most one photon"}, and that \textit{"it allows us to describe the quantum state $\rho$ of the optical field on the image plane as a mixture of a zero-photon state $\rho_0$ and a one-photon state $\rho_1$ in each time interval"}. One is thus assuming that {\it "...the probability of more than one photon arriving at the image plane is negligible"}, as stated in Ref. \cite{tsang2017}. In some other experiments, the theoretical calculations are done assuming a light source that generates single-photon quantum states, while experiments are done using intense or attenuated laser sources \cite{Boyd2019,vamivakas2018,Rehacek2016}.  In a sense, this analysis seems to be motivated by the fact even though the QFIM is an inherently quantum concept, whose terms are calculated with quantum mechanics tools, in certain cases the precision bounds can also be applied to experiments where the quantum nature of light is irrelevant \cite{Rehacek2016}.

Motivated by the previous considerations, here we will consider two types of probe beams that illuminate the phase object: $N$ copies of a multimode single-photon state and a multimode coherent state with average photon number $N$. We will demonstrate that with these types of illumination, the estimation of phase objects fulfill Eq. (\ref{matsumoto}).  We will also show that the precision bounds obtained in both cases do not coincide in general. However, if certain symmetry conditions are satisfied, the resulting QFIM is the same in both cases. As a consequence of this, our work might constitute a word of caution for experiments aimed at determining the QFIM using weak coherent states while employing single-photon quantum states, for the sake of simplicity or as an approximation, in the corresponding theoretical analysis. 

After studying the most adequate illumination scheme, a natural question arises: what kind of measurement has to be performed to attain the best precision limit?  One answer is to project the transmitted/reflected light beam in a particular set of spatial modes, a selection that depends on the specific parameter estimation problem considered \cite{tsangPRX,rehacek2017optimal}, and measure the mode amplitudes of the decomposition. One can demonstrate (see Section 1  of the Supplementary Material) that for phase objects, if the phases of the mode amplitudes do not depend on the set of parameters $\theta$, the measurement turns out to be an optimum measurement. The measurements can be done using techniques like spatial mode demultiplexing \cite{tsangPRX,tsangPRL} or evanescent coupling with different single-mode waveguides \cite{evanescent}. Although the number of modes needed varies with the different basis, and different sets of spatial modes have been considered, the number of required modes is generally large \cite{paur,rehacek2017optimal,vrehavcek2018optimal}, which renders these experiments cumbersome to implement. 

Projection onto spatial modes as a tool for highly-sensitive sensing have been considered \cite{treps2002, nathaniel1,torner,omar}. For instance, spatial mode projections were used for estimating the displacement and tilt of a Gaussian beam \cite{hsu2004,delaubert2006quantum,delaubertlimits}. In this case, selection of the real and imaginary parts of the complex value of the projection onto the same spatial mode ($TEM_{10}$) allowed to estimate the displacement and tilt, independently.  In \cite{nathaniel1} a thin nanolayer with thickness $\lambda/80$ was measured by projecting the field onto an array of spatial modes with the help of a spatial light modulator. 

In 2017 Pezze et al. \cite{pezze2017}, introduced a general procedure to generate a set of spatial modes that allows optimum parameter estimation when the condition given by Eq. (\ref{matsumoto}) is fulfilled. We apply the method described in \cite{pezze2017} to obtain the projection modes that allow optimum multiparameter estimation of phase objects with an arbitrary spatial shape. This method is a powerful tool for optimum parameter estimation since it provides, for a wide class of estimation problems, a projective measurement that saturates the QCRB. However, the proposed method is described following {\it typical} quantum estimation theory formalism, that can make difficult its comprehension for readers interested in optical sensing and metrology but not familiar with quantum estimation theory language. One of our goals is to {\it translate} to a less formal language the procedure described in \cite{pezze2017} so that readers of the physics community can find it easier to grasp the main points of the method and benefit from their results.

\section{Quantum Fisher Information matrix and the Cr\'amer-Rao lower bound}
Using the tools provided by quantum estimation theory, we aim at estimating a set of $M$ parameters ($\theta \equiv \left\{ \theta_i \right\}, i=1\dots M$) that characterize a phase object. For the sake of simplicity, we consider the state of the probe beam after reflection/transmission from the phase object to be a pure state, i.e. $\rho(\theta)=| \Phi(\theta) \rangle \langle \Phi(\theta)|$, a situation that is experimentally relevant and conveniently simplifies the calculations. The elements $F_{ij}$ of the QFIM are calculated making use of Eq. (\ref{fisher_pure}). 

One word of caution might be helpful at this point. When considering physical scenarios that involve optical phases and interferometers, one has to be careful with the selection of the quantum state that describes the experimental scheme, since apparently equal physical models can lead to different values of the QCRB. This might cause {\it interpretation problems} concerning the bounds  obtained via the QFIM  \cite{reference_phase}. Considering non-uniform phases can help to avoid some of these issues, since we can use as a reference the phase at a specific spatial location. 

\subsection*{QFIM for $N$ copies of a multimode single-photon quantum state}
We consider that the probe beam that illuminates the sample consists of $N$ independent single photons with input state $|\Phi_{in} \rangle=\int \dd x f(x) |x \rangle$, where $x$ refers to the spatial coordinate. We assume that the quantum state satisfies the normalization condition $\int \dd x |f(x)|^2=1$. After interaction with a phase object, the quantum state $\ket{\Phi}$ of the outgoing photon is
\begin{equation}
|\Phi \rangle=\int \dd x f(x) \exp \Big\{ i\,\varphi(x,\theta) \Big\} |x \rangle, \label{single_photon}
\end{equation}
where $\varphi(x)$ is the spatially-dependent phase added to the input state that depends on the set of parameters $\theta$. Making use of  Eq. (\ref{single_photon}), we obtain
\begin{equation}
\langle \Phi | \Phi_i \rangle= i\int \dd x |f(x)|^2 \left[ \frac{ \partial \varphi(x)}{\partial \theta_i} \right] 
\end{equation}
and
\begin{equation}
\langle \Phi_i | \Phi_j \rangle=\int \dd x |f(x)|^2 \left[ \frac{ \partial \varphi(x)}{\partial \theta_i} \right] \left[ \frac{\partial \varphi(x)}{\partial \theta_j} \right].
\label{matsumoto100}
\end{equation}
Note that Eq. (\ref{matsumoto100}) shows that $\langle \Phi_i | \Phi_j \rangle$ is real, so we deduce from Eq. (\ref{matsumoto}) that the QCR bound is the most informative precision bound.

We can write the elements of the QFIM  $F_{ij}^{s}$, given by Eq. (\ref{fisher_pure}), as
\begin{eqnarray}
    & &F_{ij}^{s}=4N \left\{ \int \dd x |f(x)|^2 \left[ \frac{ \partial \varphi}{\partial \theta_i} \right] \left[ \frac{\partial \varphi}{\partial \theta_j} \right] \nonumber \right. \\
    & & \left. - \left[\int \dd x |f(x)|^2 \Big( \frac{ \partial \varphi}{\partial \theta_i}  \Big) \right] \left[ \int \dd x |f(x)|^2 \Big( \frac{ \partial \varphi}{\partial \theta_j} \Big) \right] \right\}.\label{fisher1} 
\end{eqnarray}
Here, Eq. (\ref{fisher1}) corresponds to the QFIM after considering $N$ independent copies of the single-photon quantum state.

\subsection*{QFIM for a multimode coherent quantum state with mean photon number $N$}
We consider now a multimode coherent quantum state as illumination which can be defined in terms of single-mode quantum states $|\alpha_i \rangle=D(\alpha_i) |\text{vac} \rangle=\exp \left( \alpha_i \hat{a}_i^{\dagger}-\alpha_i^* \hat{a}_i \right) |\text{vac} \rangle$.
$D(\alpha_i)$ is the displacement operator and the mode normalization $\langle \alpha_i | \alpha_i \rangle=1$ holds. The multimode coherent state $\ket{\Phi}$ with $N=\sum_i |\alpha_i|^2$ mean photon number can be written as \cite{barnett} 
\begin{equation}
|\Phi\rangle=|\alpha_1 \rangle...|\alpha_N \rangle=D(\alpha_1)...D(\alpha_N) |\text{vac} \rangle.
\end{equation} 
The inner products in Eq. (\ref{fisher_pure}) have the form
\begin{equation}
\langle \Phi |  \Phi_i \rangle= \sum_{k} \langle \alpha_k | \frac{\partial \alpha_k}{\partial \theta_i}\rangle
\end{equation}
and
\begin{equation}
    \langle \Phi_i | \Phi_j \rangle = \sum_{k} \langle \frac{\partial \alpha_k}{\partial \theta_i}| \frac{\partial \alpha_k}{\partial \theta_j}\rangle + \sum_{k\ne k^{\prime}} \langle \frac{\partial \alpha_{k^\prime}}{\partial \theta_i}|\alpha_{k^\prime} \rangle \langle \alpha_k | \frac{\partial \alpha_k}{\partial \theta_j}  \rangle.
\end{equation}
Making use of the expression for the derivative of a coherent state \cite{torres2016weak}
\begin{equation}
\Big|\frac{\partial \alpha_k}{\partial \theta_i} \rangle=
i \alpha_k \left( \frac{\partial \varphi_k}{\partial \theta_i} \right) a_k^{\dagger} |\alpha_k\rangle,
\end{equation}
the inner products become
\begin{equation}
\langle \frac{\partial \alpha_k}{\partial \theta_i}|\frac{\partial \alpha_k}{\partial \theta_j} \rangle=
|\alpha_k|^2 \left( \frac{\partial \varphi_k}{\partial \theta_i} \right) \left( \frac{\partial \varphi_k}{\partial \theta_j} \right) \left( 1+|\alpha_k|^2\right)
\label{optimum_coherent}
\end{equation}
and
\begin{equation}
\langle \alpha_k |\frac{\partial \alpha_k}{\partial \theta_i} \rangle=
i |\alpha_k|^2 \left( \frac{\partial \varphi_k}{\partial \theta_i} \right).
\end{equation}
We observe again that the inner product $\langle \Phi_i |  \Phi_j \rangle$ is real for $i,j = 1, \dots , M$; which implies that the QCR bound is the most informative bound. The elements $F_{ij}^c$ of the QFIM are
\begin{equation}
\label{fisher20}
F_{ij}^c=4 \sum_k |\alpha_k|^2 \left( \frac{ \partial \varphi_k}{\partial \theta_i} \right) \left(\frac{\partial \varphi_k}{\partial \theta_j} \right).
\end{equation}
An alternative derivation of Eq. (\ref{fisher20}) is presented in Section 2 of the Supplementary Material.
For the sake of comparison with the corresponding expression for a multimode single-photon quantum state in Eq. (\ref{fisher1}), we write Eq. (\ref{fisher20}) in integral form as
\begin{equation}
\label{fisher2}
F_{ij}^c=4N \int \dd x\, |\alpha(x)|^2\, \left[ \frac{ \partial \varphi}{\partial \theta_i} \right] \left[\frac{\partial \varphi}{\partial \theta_j} \right],
\end{equation}
where $\alpha(x)$ is a normalized continuous function, i.e.,  $\int \dd x\, |\alpha(x)|^2=1$. 
Given the same spatial shape of the illumination beam, i.e., $f(x) \equiv \alpha(x)$,  the expressions for the QFIM in both cases are equal ($F_{i,j}^s=F_{i,j}^c$), if $I_{i}=0$ holds for all $i=1 \dots M$, where
\begin{equation}
I_i=\int \dd x |f(x)|^2 \Big[ \frac{ \partial \varphi}{\partial \theta_i}  \Big]. \nonumber
\end{equation}
Note that this is the case if $f(x)$ is a symmetric function, while the phase $\varphi(x,\theta) $ introduced by the object is anti-symmetric. This demonstrates that the equivalence, or non-equivalence, of the QFIM calculated using the two types of quantum states considered above, depends on the symmetry on the spatial variable $x$ of both the illumination beam and the 
acquired phase. 

\subsection*{Example: Quantum Cr\'amer-Rao bounds for the estimation of the height and the sidewall angle of a cliff-like structure.} 
\begin{figure}[t!]
    \centering
    \includegraphics[width = 8cm]{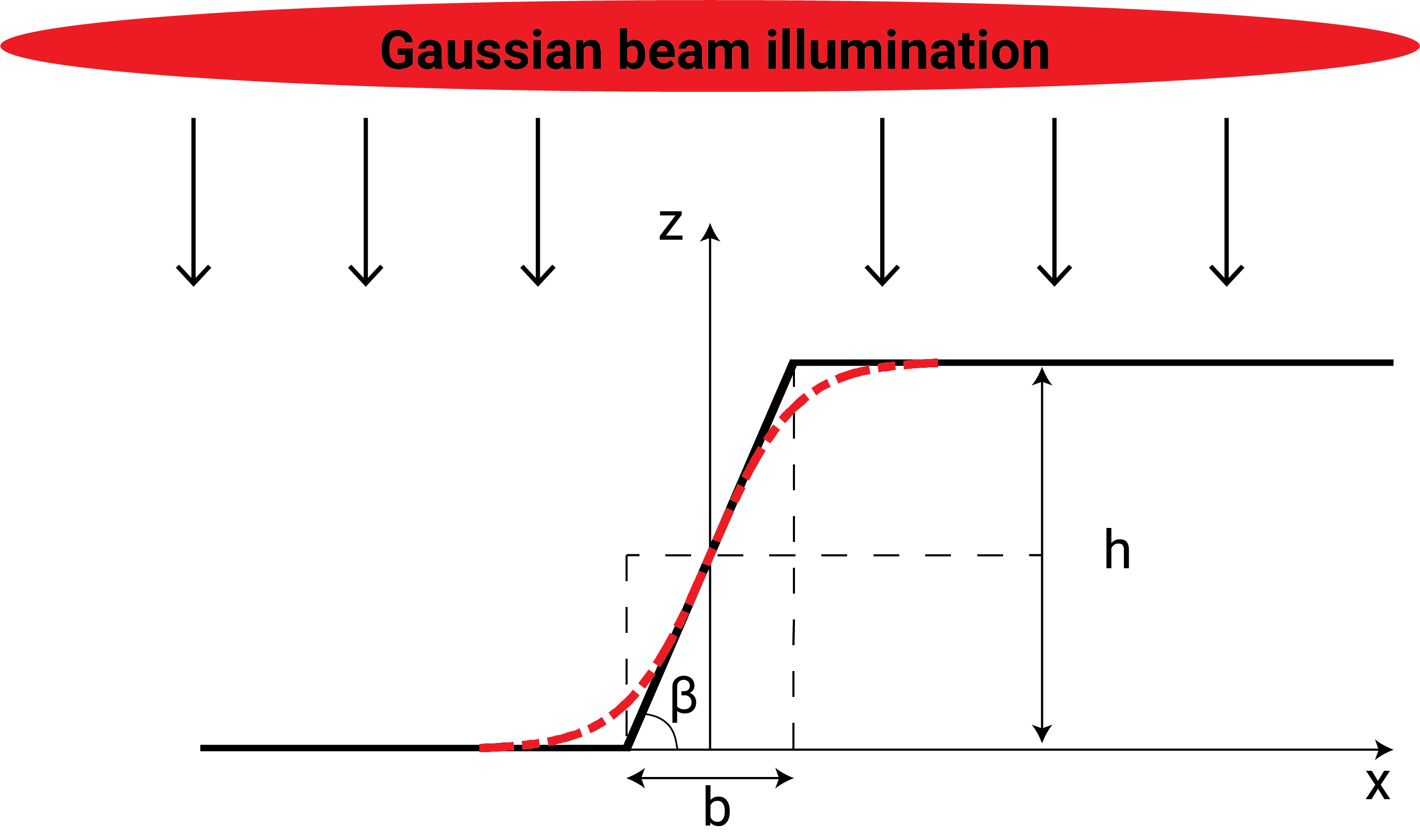}
    \caption{Sketch of a Cliff-like nano-structure with sidewall angle $\beta$ and height $h$. The red dashed line corresponds to the mathematical model of the slope. The sidewall angle $\beta$ is related to the parameter $\alpha$ as $\alpha = 2/b = 2\tan\beta /h$.}
    \label{fig:Cliff-structure}
\end{figure}

To illuminate our results,  we consider a cliff-like nanostructure made of a highly reflective material \cite{silvania1,silvania2}. This nanostructure is highly relevant in the semiconductor industry and it is characterized by two parameters: the height $h$ and the sidewall angle $\beta$, as shown in the sketch in Fig. \ref{fig:Cliff-structure}. In typical nanostructures, the height is a fraction of the wavelength of the incident light wave, and the sidewall angle is ideally close to $90^{\circ}$. However, fabrication errors produce variations around the desired values \cite{cisotto2017amplitude}.

For the sake of simplicity, we approximate the slope of the nanostructure as
\begin{equation}
S(x) = \frac{h}{2}\, \left( 1+\tanh \alpha x \right).
\end{equation}
The parameters to estimate are $\theta_1 \equiv h$ and $\theta_2 \equiv \alpha$, with $\tan \beta=\alpha h/2$. The spatial profile of the intensity of the illumination beam used for probing is assumed to be a Gaussian function
\begin{align}
    |f(x)|^2 = \left[ \frac{2}{\pi \text{w}^2} \right]^{1/2}\,\exp{-\frac{2x^2}{\text{w}^2}}.\label{Gaussian_cliff}
\end{align}
After reflection from the cliff-like structure, the optical beam acquires a spatially-dependent phase 
\begin{equation}
\label{Eq:phase_cliff}
\varphi(x,h,\alpha)=kh \left( 1-\tanh \alpha x \right).
\end{equation}
The derivatives of the phase $\varphi$ with respect to the parameters $h$ and $\alpha$ are
\begin{equation}
\frac{\partial \varphi}{\partial h}=k \left(1- \tanh \alpha x \right) \hspace{1cm} \frac{\partial \varphi}{\partial \alpha}=-kh x \sech^2 \alpha x .
\end{equation}
The elements of the QFIM for the two types of quantum states considered in this work are
\begin{eqnarray}
& & F_{11}^s=4N\, k^2 \left( 1-N_3 \right), \hspace{0.3cm} F_{11}^c \sim 4N\, k^2\left(2-N_3 \right), \label{F11} \nonumber \\
& &  F_{22}^s=F_{22}^c=4N\,h^2 N_2, \hspace{0.3cm} F_{12}^s=F_{12}^c=4N(kh)N_1,
\end{eqnarray}
where the dimensionless integrals $N_i$ are
\begin{eqnarray}
    & & N_1=k\int\,\dd x\, |f(x)|^2\,x\, \tanh (\alpha x)\, \text{sech}^2 (\alpha x) \nonumber \\
    & & N_2=k^2\int\,\dd x\, |f(x)|^2\, x^2\, \text{sech}^4 (\alpha x) \nonumber \\
    & & N_3= \int \dd x |f(x)|^2\, \sech^2 (\alpha x).
\end{eqnarray}
In Section 3 of the Supplementary Material we evaluate the dependence of the integrals $N_i$ with the parameters $w$ and $\alpha$. In most experimental implementations the beam waist of the Gaussian beam ($w$) is much larger that the size of the cliff-like structure ($\sim 1/\alpha$), so $\alpha w \gg 1$. In this case we can write to first order
\begin{eqnarray}
    & & N_1=\left[\frac{2}{\pi}\right]^{1/2} \frac{k}{w \alpha^2}, 
    \hspace{1cm} N_2=\frac{2^{1/2}\,(\pi^2-6)}{9 \pi^{1/2}} \frac{k^2}{w \alpha^3} \nonumber \\
    & & N_3=\left[\frac{8}{\pi}\right]^{1/2} \frac{1}{w \alpha}.
\end{eqnarray}
The QFIM is a real and symmetric $2\times 2$ matrix. One can deduce \cite{prussing1986} that the conditions 
\begin{eqnarray}
& & \text{Var}(h) \ge \left[ F^{-1} \right]_{11}=\frac{F_{22}}{F_{11} F_{22}-\left[ F_{12} \right]^2}, \nonumber \\ 
& & \text{Var}(\alpha) \ge \left[ F^{-1} \right]_{22}=\frac{F_{11}}{F_{11} F_{22}-\left[ F_{12}\right]^2},
\end{eqnarray}
and $F_{11}\,F_{22}-[F_{12}]^2 \ge 0$ hold. Here Var$(y)$ designates the variance of the variable $y$.

For $\alpha w \gg 1$, we have $F_{11} F_{22} \gg (F_{12})^2$. The relative precision error $\sigma_{\alpha} \equiv \sqrt{Var(\alpha)}/\alpha$ of the estimation of the parameter $\alpha$ is
\begin{equation}
\sigma_{\alpha} \geq \frac{0.85}{kh}\,\sqrt{\frac{\alpha w}{N}}\label{sigma_alpha}
\end{equation}
for both types of multimode quantum states. 
On the other hand, for the estimation of the height $h$, the relative precision error $\sigma_{h} \equiv \sqrt{Var(h)}/h$ for multimode single-photon quantum states is
\begin{equation}
\sigma_h \geq \frac{1}{2kh}\,\sqrt{\frac{1}{N}}\label{sigma_h_sp}
\end{equation}
whereas for a multimode coherent quantum state is
\begin{equation}
\sigma_h \geq \frac{1}{2\sqrt{2}\,kh}\,\sqrt{\frac{1}{N}}.\label{sigma_h_cs}
\end{equation}
The estimation of the parameter $\alpha$ with multimode coherent and single-photon quantum states gives the same result because the derivative of the phase given by Eq. (\ref{Eq:phase_cliff}) with respect to the parameter $\alpha$ is an antisymmetric function of the spatial coordinate ($\sim x \sech \alpha x$) such that $I_1=0$. This is not the case for the derivative of the phase with respect to the parameter $h$ ($I_2 \neq 0$), hence the precision bounds given by the corresponding elements of the inverse QFIM for multimode coherent and single-photon quantum states lead to different bounds for the estimation of the height $h$, implying that a coherent state with average photon number $N$ allows better estimation precision than $N$ copies of a single-photon state. 

\section{Optimum estimation measurement by projection onto spatial modes}
\subsection*{General scheme}
For the sake of simplicity, let us consider the case of two-parameter phase estimation. The generalization of the method for the estimation of more than two parameters is straightforward. The probe light beam consists of a multimode single-photon pure state with spatial shape $f(x)$. After reflection/transmission, the sample adds a spatially-varying phase $\varphi(x,\theta_1,\theta_2)$ that depends on two parameters: $\theta_1$ and $\theta_2$. We can thus write the quantum state of the output light beam as
\begin{equation}
\label{qstate}
|\Phi \rangle=\int \dd x \,f(x) \exp \Big[i \varphi(x,\theta_1,\theta_2) \Big] a^{\dagger}(x) |0\rangle
\end{equation}
where $x$ designates the spatial coordinate and $a^{\dagger}(x)$ is the creation operator.  As starting point, we consider two values of the parameters as reference: $\theta_{10}$ and $\theta_{20}$. We aim at estimating the values of $\Delta \theta_1$ and $\Delta \theta_2$ where $\theta_1=\theta_{10}+\Delta \theta_1$ and $\theta_2=\theta_{20}+\Delta \theta_2$. The parameter differences
$\Delta \theta_1$ and $\Delta \theta_2$ are assumed to be small. We define the reference state as
\begin{equation}
\label{phi0}
|\Phi_0 \rangle=\int \dd x f(x) \exp \Big[i \varphi_0 \Big] |x\rangle
\end{equation}
where $\varphi_0 \equiv \varphi(x,\theta_{10},\theta_{20})$ and $|x \rangle \equiv a^{\dagger}(x)|0 \rangle$. Following \cite{pezze2017,villegas2020spatial} we construct two quantum states orthogonal to $|\Phi_0 \rangle$,
\begin{eqnarray}
    & & \ket{\omega_1} = \ket{\Phi_1} + \langle \Phi_1\ket{\Phi_0}\ket{\Phi_0}\label{omega1}\\
    & & \ket{\omega_2} = \ket{\Phi_2} + \langle \Phi_2\ket{\Phi_0}\ket{\Phi_0}.\label{omega2}
\end{eqnarray}
All derivatives are taken at $\theta_1=\theta_{10}$ and $\theta_2=\theta_{20}$. We  generate an orthonormal basis of quantum states obtained through the Gramm-Schmidt process
\begin{eqnarray}
& & \ket{\gamma_0}=\ket{\Phi_0}  \label{gamma10}\\  
& & \ket{\gamma_1} = \frac{1}{\Omega_{11}^{1/2}} \ket{\omega_1} \label{gamma11}\\
& & \ket{\gamma_2} = \Big( \frac{\Omega_{11}}{\det\, \Omega}\Big)^{1/2}\, \Big[ \ket{\omega_2} - \frac{\Omega_{12}}{\Omega_{11}} \ket{\omega_1} \Big].\label{gamma12}
\end{eqnarray}
In what follows, we demonstrate that for small values of $\Delta \theta_1$ and $\Delta \theta_2$, projecting the quantum state in Eq. (\ref{qstate}) onto the set of spatial modes $\{\ket{\gamma_k}\}$, given by Eqs. (\ref{gamma10})-(\ref{gamma12}), provides an optimum estimation of the parameters $\theta_1$ and $\theta_2$. We put forward two examples that will help clarifying the main characteristics of the method described above.

\subsection*{Example I: optimum estimation of the height of a cliff-like structure}
\begin{figure*}[t!]
    \centering
    \includegraphics[width = \textwidth]{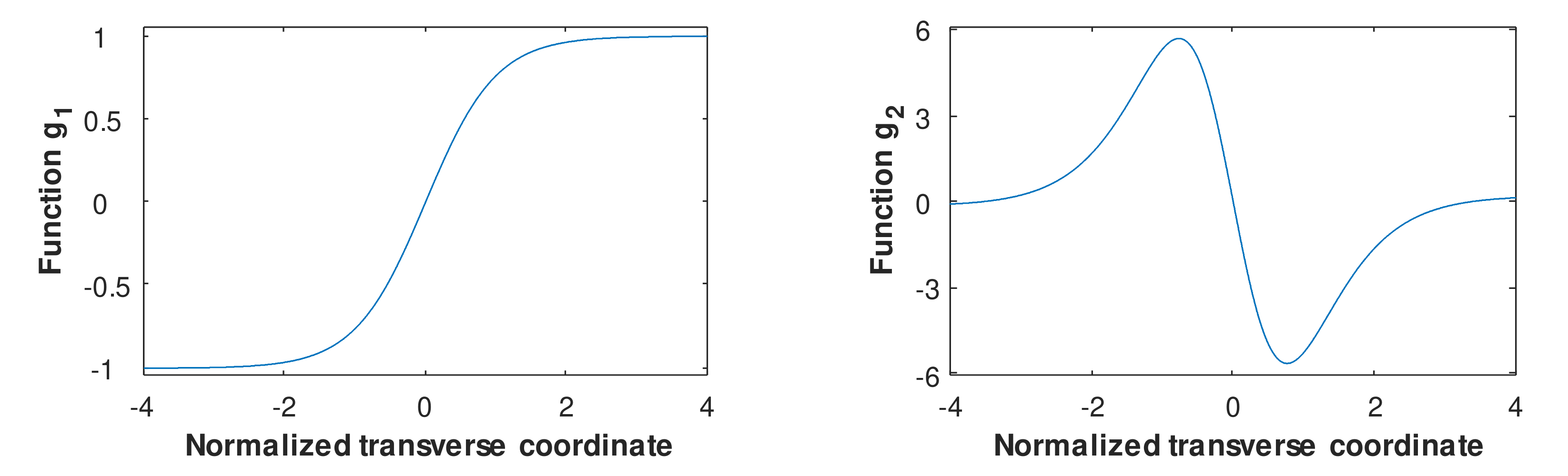}
    \caption{Modulus of the functions $g_1(y)$ and $g_2(y)$. The normalized transverse coordinate is $y=\alpha_0 x$, where $x$ is the spatial coordinate $x$ and $\alpha_0=(2 \tan \beta_0)/h_0$. The wavelength is $\lambda=633$ nm, the reference height is $h_0=\lambda/4$ and the sidewall angle is $\beta_0=80^{\circ}$.}
    \label{figuresg1g2}
\end{figure*}

We consider as first example a case of single-parameter estimation. We assume that the side-wall  angle parameter $\alpha$ has a fixed value $\alpha=\alpha_0$, and we aim at estimating the height of the cliff-like structure, i.e., the parameter $\theta_1=h$.  The reference value of the height is $h_0$, so we want to estimate $\Delta h$ where $h=h_0+\Delta h$. The spatially-dependent phase added by the sample in reflection is $\varphi(x)=k h \left( 1-\tanh \alpha_0 x \right)$, so that the quantum state after reflection from the cliff-like structure is $|\Phi \rangle= \int \dd x f(x) \exp \left(i \varphi \right)|x\rangle$. The derivative of the phase at $h=h_0$ is 
\begin{equation}
\frac{\partial \varphi}{\partial h}\bigg|_{h=h_0}=k \left[1- \tanh \alpha_0 x \right].
\end{equation}
The derivative of the quantum state $|\Phi \rangle$ is
\begin{equation}
|\Phi_1 \rangle= i k \int \dd x f(x) \Big[ 1-\tanh (\alpha_0 x) \Big]\, \exp \left[ i\varphi_0 \right]  |x\rangle, 
\label{Phi_h}
\end{equation}
where $\varphi_0= k h_0 \big[ 1-\tanh (\alpha_0 x) \big]$. Making use of Eqs. (\ref{phi0}) and (\ref{Phi_h}), the inner products are
\begin{eqnarray}
    & & \langle \Phi_0|\Phi_1 \rangle=\int \dd x |f(x)|^2 \left( \frac{\partial \varphi}{\partial h}\right)=ik  \nonumber \\
    & & \label{inner1} \langle \Phi_1|\Phi_1 \rangle=\int \dd x |f(x)|^2 \left( \frac{\partial \varphi}{\partial h}\right)^2=k^2(2-N_3).
\end{eqnarray}
We have assumed the experimentally relevant case in which the spatial dimensions of the illumination field are much larger than the spatial features of the cliff-like nanostructure, i.e., the Gaussian function is assumed to be constant inside the integrals. This approximation is valid if $w \alpha_0 \gg 1$ (see Section 3 of the Supplementary Material). Making use of $\Omega_{11}= k^2 (1-N_3)$, the quantum state $|\gamma_1\rangle$ is
\begin{equation}
\label{gamma1}
|\gamma_1 \rangle=\int \dd x \,f(x)\,g_1(x)\, \exp (i \varphi_0) |x\rangle,
\end{equation}
where
\begin{equation}
\label{functiong1}
g_1(x)=- \frac{i}{ (1-N_3)^{1/2}}\,\tanh (\alpha_0 x).
\end{equation}
Figure \ref{figuresg1g2} (a) shows the modulus of the function $g_1$ for a typical value of the parameters $h_0$ and $\alpha_0$. 

In order to obtain an analytical expression of the value of the mode projection of $|\Phi \rangle$ onto the quantum modes $\ket{\gamma_0}$ and $\ket{\gamma_1}$, we make use of the Taylor expansion of $\exp[i(\varphi-\varphi_0)]$ to second order in $\Delta h$. If we define the mode detection probabilities as $p_0=|\langle \gamma_0 |\Phi \rangle|^2$ and $p_1=|\langle \gamma_1 |\Phi \rangle|^2$, we obtain (see Section 4 in the Supplementary Material) that to second order in $\Delta h$ 
\begin{eqnarray}
    & & p_0=1- (1-N_3)\,(k\Delta h)^2,  \nonumber \\
    & &  \label{projection1} p_1=(1-N_3) (k \Delta h)^2.
\end{eqnarray}

We can see that to second order in $\Delta h$, we obtain $p_0+p_1=1$. Because of this, we can consider the estimation of the parameter $h$ by projecting onto the spatial mode $|\gamma_1 \rangle$ as a Bernoulli process with variance $p_1 (1-p_1)$ \cite{milburn}.  The probability to detect a reflected/transmitted photon under ideal detection efficiency is $p_1$. The sensitivity of the estimation of $h$ is given by:
\begin{equation}
    \text{Var} (h) =\frac{p_1(1-p_1)}{\left( \partial p_1/\partial \Delta h \right)^2}=\frac{1}{4k^2\,(1-N_3)} .
\end{equation}
Comparison with Eq. (\ref{F11}) shows that $\text{Var} (h)$ can be written as $\text{Var} (h)=1/F_{11}$, that demonstrates that the projection onto $| \gamma_1 \rangle$ is a measurement that saturates the value of precision given by the quantum CRB.

One final consideration can be illuminating at this point. One might wonder why not to use as projector the quantum state $|\Phi_1\rangle$?
In \cite{pezze2017} they demonstrate that for projectors that are NOT orthogonal to $|\Phi_0\rangle$, as it is the case of using $|\Phi_1\rangle$, the condition to saturate the CRB (see Eq. (8) in \cite{pezze2017} for one parameter estimation) is
\begin{equation}
\label{second}
\Im \Big[ \langle \Phi_1 |\Phi_1 \rangle \langle \Phi_1 |\Phi_0 \rangle \Big]=|\langle \Phi_0 |\Phi_1 \rangle|^2 \Im \Big[  \langle \Phi_1 |\Phi_0 \rangle \Big].
\end{equation}
Substituting Eq. (\ref{inner1}) in Eq. (\ref{second}), we obtain $k^3(2-N_3) \ne k^3$, which allows us to conclude that the use of $|\Phi_1 \rangle$ as projector does not provide an optimum estimation of the height $h$. In \cite{delaubertARX,delaubertlimits},  the authors considered the estimation of the center $x_0$ of a Gaussian beam. In this case, one finds that $\langle \Phi_0 |\Phi_1 \rangle=0$, they are orthogonal, so $|\gamma_1\rangle =|\Phi_1 \rangle$, differently from the case of the estimation of the height $h$ of a cliff-like structure considered above.

\subsection*{Example II: estimation of the parameters $h$ and $\alpha$ of a cliff-like structure}
We consider now the case where we aim at estimating both the height $h$ and the steepness of the cliff, represented by the parameter $\alpha$ in our analytical model of the slope. We define $\theta =(\theta_1,\theta_2) = (h,\alpha)$. This corresponds to a two-parameter estimation. Similarly to the one-parameter case, we consider reference values $h_0$ and $\alpha_0$, and small variations of the parameters around the reference values: $h=h_0+\Delta h$ and $\alpha=\alpha_0+\Delta \alpha$.

The quantum state $|\gamma_1 \rangle$ is still given by Eq. (\ref{gamma1}), and the inner products involving $|\Phi_1 \rangle$ are given by Eq. (\ref{inner1}). The inner products that involve $|\Phi_2 \rangle$ are
\begin{eqnarray}
    & & \bra{\Phi_2}\Phi_0\rangle = 0 \nonumber \\
    & &  \label{inner2} \bra{\Phi_1}\Phi_2\rangle=k h_0 N_1 \\
    & & \bra{\Phi_2}\Phi_2\rangle=h_0^2 N_2. \nonumber
\end{eqnarray}
The quantum state $|\gamma_2 \rangle$ reads
\begin{align}
    \ket{\gamma_2} =\int\,\dd x\,f(x)\,g_2(x)\, \exp(i \varphi_0) |x\rangle.
\end{align}
where
\begin{equation}
\begin{split}
    g_2(x) & =i\Big( \frac{\Omega_{11}}{\det\, \Omega} \Big)^{1/2}\,\times \\
    & \Big\{-k h_0 \,x\,\text{sech}^2 (\alpha_0 x)+k \frac{\Omega_{12}}{\Omega_{11}}\,\tanh(\alpha_0 x)  \Big\},
\end{split}
\end{equation}
$\Omega_{11}=k^2(1-N_3)$, $\Omega_{22}=h_0^2 N_2$ and $\Omega_{12}=kh_0 N_1$. Figure \ref{figuresg1g2}(b) shows the shape of function $g_2$ for typical values of the parameters $h_0$ and $\alpha_0$. 

Similar to the one-parameter estimation case considered above, in order to obtain an analytical result we make use of the Taylor expansion of the exponential term to second order in $\Delta h$ and $\Delta \alpha$ (see Section 5 in the Supplementary Material). The mode detection probabilities $p_0=|\langle \gamma_0 |\Phi \rangle|^2$, $p_1=|\langle \gamma_1 |\Phi \rangle|^2$ and $p_2=|\langle \gamma_2 |\Phi \rangle|^2$, up to second order on $\Delta h$ and $\Delta \alpha$, read
\begin{eqnarray}
    & & p_0=1-k^2(2-N_3) (\Delta h)^2-N_2 (h_0 \Delta \alpha)^2 
    \nonumber \\
    & & \hspace{1cm}
    -2\, k\, h_0 N_1 (\Delta h)(\Delta \alpha) +(k \Delta h)^2,  \\
    & & p_1=k^2\,(1-N_3) (\Delta h)^2+\frac{N_1^2}{1-N_3}\, (h_0 \Delta \alpha)^2 ,
    \nonumber \\
    & & \hspace{1cm}
    +2\, k \, h_0 \, N_1 (\Delta h)\,(\Delta \alpha) \\
    & & p_2=\left( N_2-\frac{N_1^2}{1-N_3} \right) (h_0 \Delta  \alpha)^2 .   
\end{eqnarray}
One can easily verify that $p_0+p_1+p_2=1$. In Section 6 of the Supplementary Material we show that the classical and quantum Fisher information matrix coincide, which demonstrates that projection onto the quantum states $|\gamma_1 \rangle$ and $|\gamma_2 \rangle$ is an optimum measurement that saturates the QCRB.

\section{Conclusions}
We have demonstrated that the most informative bound for the precision in the estimation of a set of parameters that characterize a shaped phase object is the Quantum  Cr\'amer Rao bound. To derive this bound we have calculated the QFIM for two types of light waves, namely $N$ copies of a multimode single-photon quantum state and a multimode coherent quantum state with mean photon number $N$. We have shown that the equivalence of these quantum states for parameter estimation of phase objects depends on the spatial symmetry of the phase introduced by the phase object.  The results presented in this work are a word of caution for experiments measuring the QFIM that make use of weak coherent states, while use single-photon quantum states in the corresponding theoretical analysis for the sake of simplicity. 

We have also shown a measurement scheme for multiparameter estimation of features of phase objects, that allows the estimation with the best precision allowed by the light-matter interaction considered. In order words, the measurement scheme saturates the quantum Cr\'amer Rao precision bound. The method can be described as \textit{spatial spectroscopy}, since it mimics in the spatial domain what conventional spectroscopy methods do in the (temporal) frequency domain (hyperspectral imaging). The technique, that follows the general guidelines put forward in \cite{pezze2017}, consists in projecting the light reflected or transmitted from the phase object onto special modes with engineered spatial properties. The required number of spatial modes needed to estimate a set of parameters is the same as the number of parameters. This is a striking contrast with other techniques that require the use of a large number of spatial modes to achieve good precision in the estimation of the unknown parameters. 

We have considered two examples to help clarifying the main characteristics of the method, with the aim of making {\it easier} its understanding to  readers of the physics community who are interested in high precision optical sensing and imaging, but are not familiar with the formal language of quantum estimation theory.

For the experimental implementation, one can think of using a spatial light modulator where the projection modes can be imprinted. A similar scheme was implemented in \cite{nathaniel1} for estimating a step height smaller than $10$ nm, i.e., one eightieth ($1/80$) of the wavelength used with a standard error in the picometer scale. One can also consider measuring the wavefront of the output light beam (amplitude and phase) with any experimental method available, such as Ptychography \cite{ptycography2008}, digital holography \cite{roadmap2021}, and project computationally the optical field retrieved experimentally onto the optimum spatial modes that provides the best spatial resolution.

\appendix
\section{Supplementary Material}

\subsection{Conditions for a measurement based on spatial mode projection to saturate the Cr\'amer-Rao precision bound.}
It has been shown that one can select a set of spatial modes to estimate with optimum precision a single parameter \cite{hsu2004,tsangPRX,rehacek2017optimal}, i.e., a measurement that saturates the Cr\'amer-Rao precision bound. Here we show what conditions should satisfy a basis of spatial modes for optimum multiparameter estimation of a shaped phase object when the illumination probe is a multimode single-photon quantum state with spatial shape $f(x)$.

A shaped phase object generates a phase shift of the reflected/transmitted photons $\varphi(x, \theta)$, i.e, the output quantum state is
\begin{equation}
|\Phi \rangle=\int \dd x\, f(x) \exp \left[ i\varphi(x,\theta) \right]\, a^{\dagger}(x) |0 \rangle,
\end{equation}
where $\theta \equiv \left\{\theta_1,\theta_2 \hdots \theta_M \right\}$ is the set of parameters to estimate. The inner products read
\begin{equation}
\langle \Phi_i|\Phi_j \rangle=\int \dd x\, |f(x)|^2 \left( \frac{\partial \varphi}{\partial \theta_i} \right)\,\left( \frac{\partial \varphi}{\partial \theta_j} \right) 
\end{equation}
and
\begin{equation}
\langle \Phi|\Phi_i \rangle=\int \dd x\, |f(x)|^2 \left( \frac{\partial \varphi}{\partial \theta_i} \right).
\end{equation}
The elements of the QFIM are $F_{ij}=4 \Omega_{ij}$, where
\begin{equation}
\Omega_{ij}=\langle \Phi_i|\Phi_j \rangle+ \langle \Phi_i|\Phi_0 \rangle \langle \Phi_j|\Phi_0 \rangle.
\end{equation}
The function that determines the spatial shape of the output quantum state, $f(x) \exp \left[i \varphi(x,\theta) \right]$, can be expanded into a basis of spatial modes as
\begin{equation}
f(x) \exp \left[ i \varphi(x,\theta) \right]=\sum_n C_n(\theta) |u_n \rangle,
\end{equation}
where $|u_n \rangle$ are the elements of the basis and $C_n$ are the complex mode amplitudes that depend on the values of the set $\theta$. We can write
\begin{equation}
\langle \Phi_i|\Phi_j \rangle=\sum_{n} \left( \frac{\partial C_n^*}{\partial \theta_i} \right)\,\left( \frac{\partial C_n}{\partial \theta_j} \right)
\end{equation}
and
\begin{equation}
\langle \Phi|\Phi_i \rangle=\sum_{n} C_n^*\, \left( \frac{\partial C_n}{\partial \theta_i} \right).
\end{equation}
If we write $C_n=\rho_n \exp (i \varphi_n)$, with $\rho_n$ being the modulus and $\varphi_n$ the phase of the mode amplitudes, the elements $\Omega_{ij}$ can be written as
\begin{eqnarray}
& & \Omega_{ij}=\sum_n \left\{ \left( \frac{\partial \rho_n}{\partial \theta_i} \right)\, \left( \frac{\partial \rho_n}{\partial \theta_j} \right)+
\rho_n^2\,\left( \frac{\partial \varphi_n}{\partial \theta_i} \right)\, \left( \frac{\partial \varphi_n}{\partial \theta_j} \right) \right.  \left. \right.\nonumber\\
& & \left.-\left[ \rho_n^2\,\left( \frac{\partial \varphi_n}{\partial \theta_i} \right) \right]\,\left[ \rho_n^2\,\left( \frac{\partial \varphi_n}{\partial \theta_j} \right) \right] \right\}.\label{demo10}
\end{eqnarray}
In the derivation of Eq. (\ref{demo10}), we have used that for phase objects, one has $\Im \langle \Phi_i|\Phi_j \rangle=0$, which implies that
\begin{equation}
\rho_n\,\left( \frac{\partial \rho_n}{\partial \theta_i} \right)\, \left( \frac{\partial \varphi_n}{\partial \theta_j} \right)=\rho_n\,\left( \frac{\partial \rho_n}{\partial \theta_j} \right)\, \left( \frac{\partial \varphi_n}{\partial \theta_i} \right).
\end{equation}
If we consider an experimental scheme based on the projective measurement onto the modes $|u_n \rangle$, the elements $F_{ij}^C$ of the classical Fisher information matrix elements are
\begin{equation}
F_{ij}^C=\sum_n \frac{1}{p_n}\,\left( \frac{\partial p_n}{\partial \theta_i} \right)\, \left( \frac{\partial p_n}{\partial \theta_j} \right).
\end{equation}
Using the fact that the probabilities are $p_n=\rho_n^2$, one can easily see that 
\begin{equation}
\label{classicaldemo10}
F_{ij}^C=\sum_n \rho_n\,\left( \frac{\partial \rho_n}{\partial \theta_i} \right)\, \left( \frac{\partial \rho_n}{\partial \theta_j} \right).
\end{equation}
Comparing Eqs. (\ref{demo10}) and (\ref{classicaldemo10}) we see that selecting a base $|u_n \rangle$ where for the whole set of parameters $\theta$, one has $\partial \varphi_n/\partial \theta_i=0$, the measurement constitutes an optimum measurement, i.e., the quantum and classical Fisher information matrix coincide. 

Let us consider two examples. First, it has been demonstrated \cite{tsangPRX} that the estimation of the distance $d$ between two incoherent optical point sources can be estimated with optimum precision if one uses a set of Hermite-Gauss (HG) modes. Secondly, in \cite{hsu2004} it was demonstrated that the position of a Gaussian beam can be estimated optimally by projecting the displaced Gaussian beam into the set of HG modes. The mode amplitudes of the projection are found to be 
\begin{equation}
C_n=\frac{1}{\sqrt{n!}}\,\left( \frac{d}{w_0} \right)^n\, \exp\left(- \frac{d^2}{2w_0^2} \right)
\end{equation}
where $n$ is the order of the Hermite-Gauss mode and $w_0$ is a beam width that characterize the HG modes. Notice that these coefficients do not contain a phase that depends on $d$.

\subsection{Alternative derivation of Eq. (13) of the main text}
The introduction of a spatially dependent phase $\left\{ \varphi_k \right\}$ for each spatial coordinate (index $k$) is an unitary operation that can be represented by the operator 
\begin{equation}
 U=\exp \Big[ i \sum_k \varphi_k(\theta)\, a_k^{\dagger} a_k \Big],
\end{equation}
such that the output quantum state is
\begin{equation}
 |\Phi(\theta)\rangle=U(\theta) | \alpha \rangle.
\end{equation}
The derivative of the quantum state with respect parameter $\theta_i$ is
\begin{equation}
 |\Phi_i \rangle=\left( \frac{\partial U}{\partial \theta_i}\right) |\alpha \rangle=i\sum_k \left( \frac{\partial \varphi_k}{\partial \theta_i} \right) a_k^{\dagger} a_k\, | \alpha \rangle 
\end{equation}
so that 
\begin{equation}
\label{ap1}
 \langle \Phi|\Phi_i \rangle=i\sum_k \left( \frac{\partial \varphi_k}{\partial \theta_i} \right) \langle \alpha |a_k^{\dagger}a_k|\alpha \rangle .
\end{equation}
Similarly we can write
\begin{eqnarray}
 & & \langle \Phi_i|\Phi_j\rangle=\sum_{k}\,\sum_{k^{\prime}}\, \left( \frac{\partial \varphi_k}{\partial \theta_i} \right) \left( \frac{\partial \varphi_{k^{\prime}}}{\partial \theta_j} \right) \, \langle \alpha|a_k^{\dagger} a_{k}\,a_{k^{\prime}}^{\dagger} a_{k^{\prime}} | \alpha \rangle \nonumber \\
 & & =\sum_{k}\, \left( \frac{\partial \varphi_k}{\partial \theta_i} \right) \left( \frac{\partial \varphi_{k}}{\partial \theta_j} \right) \, \langle \alpha|a_k^{\dagger} a_{k}\,a_{k}^{\dagger} a_{k} | \alpha \rangle \label{ap2} \\
 & & +\sum_{k \ne k^{\prime}}\, \left( \frac{\partial \varphi_k}{\partial \theta_i} \right) \left( \frac{\partial \varphi_{k^{\prime}}}{\partial \theta_j} \right) \, \langle \alpha|a_k^{\dagger} a_{k} |\alpha \rangle \langle a_{k^{\prime}}^{\dagger} a_{k^{\prime}} | \alpha \rangle \nonumber
\end{eqnarray}
The elements of the QFIM are 
\begin{eqnarray}
 & & F_{ij}=4\langle \Phi_i|\Phi_j\rangle-4\langle \Phi|\Phi_i\rangle\,\langle \Phi|\Phi_j \rangle \nonumber \\
 & & =4 \sum_{k}\,\left( \frac{\partial \varphi_k}{\partial \theta_i} \right) \left( \frac{\partial \varphi_{k}}{\partial \theta_j} \right) \Big\{ \langle \alpha|a_k^{\dagger} a_{k}\,a_{k}^{\dagger} a_{k} | \alpha \rangle-\Big[ \langle \alpha|a_k^{\dagger} a_{k} | \alpha \rangle \Big]^2 \Big\} \label{ap3} \nonumber \\
 & & =\sum_{k}\, \left( \frac{\partial \varphi_k}{\partial \theta_i} \right) \left( \frac{\partial \varphi_{k}}{\partial \theta_j} \right) \langle \left( \Delta N_k \right)^2 \rangle, 
\end{eqnarray}
where $N_k \equiv \langle a_k^{\dagger} a_k \rangle$ and the variance is $\langle \left( \Delta N_k \right)^2 \rangle= \langle N_k^2  \rangle-\langle N_k \rangle^2$. Making use of Eq. (\ref{ap3}) and the fact that $\langle \left( \Delta N_k \right)^2 \rangle=|\alpha_k|^2$ for quantum coherent states, we obtain Eq. (13) on the main text.

\subsection{Validity of constant field approximation for integration}
In general, all integrals $N_i$ are of the form
\begin{equation}
N_i=k^m\,\int \dd x\ |f(x)|^2\, x^m\,\tanh^n (\alpha_0 x) \, \text{sech}^p (\alpha x) 
\end{equation}
where the integers $m$, $n$ and $p$ varies for different integrals. If we expand the Gaussian function as
\begin{equation}
\Big( \frac{2}{\pi\text{w}^2} \Big)^{1/2} \exp \Big[ -\frac{2x^2}{\text{w}^2} \Big] \sim  \Big( \frac{2}{\pi\text{w}^2} \Big)^{1/2} \left[ 1- \frac{2 x^2}{\text{w}^2} \right]
\end{equation}
and define $y=\alpha x$, we obtain that
\begin{eqnarray}
& & N_i= \Big( \frac{2}{\pi} \Big)^{1/2}\frac{k^m}{\alpha^{m+1} \text{w}}\, \int \dd y\,y^m\,\tanh^n (y) \, \text{sech}^p (y) \nonumber \\
& & \times \left\{  1-\frac{2}{(\text{w} \alpha)^2}\,\frac{\int \dd y\,y^{m+2}\,\tanh^n (y) \, \text{sech}^p (y) }{\int \dd y\,y^m\,\tanh^n (y) \, \text{sech}^p (y)} \right\}.
\end{eqnarray}
First, notice that to first order, the dependence of $N_i$ on the parameters w and $\alpha$ is $N_i \sim 1/[\text{w}\alpha^{m+1}]$. Second, the approximation of neglecting the spatial dependence of the Gaussian functions is valid if $1/(\text{w} \alpha)^2 \ll 1$.

\subsection{Calculation of the analytical approximations of the inner products for Example I of the main text}
Here we calculate the mode projections of the quantum state 
\begin{equation}
|\Phi \rangle=\int \dd x\, f(x) \exp(i \varphi) a^{\dagger}(x) | 0 \rangle,
\end{equation}
where
\begin{equation}
    f(x) = \Big( \frac{2}{\pi\text{w}^2} \Big)^{1/4} \exp \Big[ -\frac{x^2}{\text{w}^2} \Big],
\end{equation}
and 
\begin{equation}
\varphi=kh \left[ 1-\tanh(\alpha x) \right].
\end{equation}
The basis of quantum states has elements
\begin{equation}
|\gamma_k  \rangle=\int \dd x\, g_k(x) \exp (i \varphi_0)\, a^{\dagger}(x) | 0 \rangle,
\end{equation}
where $g_k(x)$ is a function characteristic of  each mode and
\begin{equation}
\varphi_0 =kh_0 \left[ 1-\tanh(\alpha_0 x) \right].
\end{equation}
The expression of the mode projections reads in general:
\begin{equation}
\label{inner10}
\langle \gamma_k|\Phi  \rangle=\int \dd x\, g_k^*(x)\, f(x)\, \exp \left[ i (\varphi- \varphi_0) \right].
\end{equation}
In order to obtain analytical approximations to second-order in $\Delta h$ , we make use of the Taylor expansion
\begin{equation}
\label{taylor1}
    \exp \left[ i (\varphi-\varphi_0) \right]= 1-\frac{A^2}{2} (k\Delta h)^2 +i \,A(k \Delta h), 
\end{equation}
where
\begin{equation}
\label{A1}
    A=1-\tanh (\alpha_0 x). 
\end{equation}
Making use of Eqs. (\ref{taylor1}) and (\ref{A1}), that $g_0(x)=f(x)$ and
\begin{equation}
\label{gamma10}
g_1 (x)=- \frac{i}{ (2-N_3)^{1/2}}\,f(x)\,\tanh (\alpha_0 x), 
\end{equation}
we obtain for Example I:
\begin{eqnarray}
    & & \langle \gamma_0 |\Phi \rangle =1-\frac{(k\Delta h)^2}{2}\, (2-N_3) + i k \Delta h, \nonumber \\
    & &  \label{projection1} \langle \gamma_1 |\Phi \rangle =(1-N_3)^{1/2} \Big[ k \Delta h +i\, (k\Delta h)^2 \Big] ,
\end{eqnarray}
where
\begin{equation}
N_3= \int \dd x |f(x)|^2\, \sech^2 (\alpha_0 x).
\end{equation}

\subsection{Calculation of the analytical approximations of the inner products for Example II of the main text}
For the case of the two-parameter estimation of the parameters $h$ and $\alpha$ (Example II in the main text), the functions $g_0(x)$ and $g_1(x)$ are the same than in the one-parameter estimation scenario. The function $g_2(x)$ is
\begin{equation}
\begin{split}
    g_2(x) & = i\Big( \frac{\Omega_{11}}{\det\, \Omega} \Big)^{1/2}\,f(x)  \times \\
    & \Big\{-k h_0 \,x\,\text{sech}^2 (\alpha_0 x) +k \frac{\Omega_{12}}{\Omega_{11}}\,\tanh(\alpha_0 x)  \Big\} 
\end{split}
\end{equation}
where $\Omega_{11}=k^2(1-N_3)$, $\Omega_{22}=h_0^2 N_2$, $\Omega_{12}=kh_0 N_1$. In order to obtain analytical approximations to second-order in $\Delta h$ and $\Delta \alpha$, we make use of the Taylor expansion
\begin{eqnarray}
    & & \label{taylor2} \exp \left[ i (\varphi-\varphi_0) \right]= \Big[ 1-\frac{A^2}{2} (k\Delta h)^2 -\frac{B^2}{2} (h_0 \Delta \alpha)^2 \nonumber \\
    & & + A B (k \Delta h)\,(h_0 \Delta \alpha) \Big] +i \Big[ A(k \Delta h) -B (h_0 \Delta \alpha) \nonumber \\
    & & +C\frac{h_0}{k}\,(\Delta \alpha)^2 -B(\Delta h)\,(\Delta \alpha) \Big], 
\end{eqnarray}
where
\begin{eqnarray}
    & & A=1-\tanh (\alpha_0 x), \nonumber  \\
    & & \label{ABC} B=k x \sech^2 (\alpha_0 x), \\
    & & C = k^2 x^2 \sech^2(\alpha_0 x)\,\tanh{(\alpha_0 x}). \nonumber
\end{eqnarray}
Making use of Eqs. (\ref{taylor2}) and (\ref{ABC}), the mode projections are 
\begin{eqnarray}
    & & \langle \gamma_0|\Phi \rangle =1-(2-N_3)\, \frac{(k \Delta h)^2}{2}-N_2 \frac{(h_0 \Delta \alpha)^2}{2} \\
    & & \hspace{1.2cm}-N_1 (k \Delta h)(h_0 \Delta \alpha) +i \Big[ (k \Delta h)+N_4 \frac{h_0}{k} (\Delta \alpha)^2 \Big],\nonumber
\end{eqnarray}
\begin{eqnarray}
    & & \langle \gamma_1|\Phi \rangle =(1-N_3)^{1/2}\,(k\Delta h)+\frac{N_1}{(1-N_3)^{1/2}} (h_0 \Delta \alpha) \nonumber \\
    & &  \hspace{1.2cm}\frac{N_6}{(1-N_3)^{1/2}} \frac{h_0}{k} (\Delta \alpha)^2 +\frac{N_1}{(1-N_3)^{1/2}} (\Delta h)\, (\Delta \alpha) \nonumber\\
    & & \hspace{1.2cm}+i \; \Big[ (1-N_3)^{1/2} (k \Delta h)^2 \Big.\nonumber\\
    & & \Big.\hspace{1.2cm} + \frac{N_1}{(1-N_3)^{1/2}} (k \Delta h)\, (h_0 \Delta \alpha) \Big],
\end{eqnarray}
and
\begin{eqnarray}
    & &  \langle \gamma_2|\Phi \rangle=\left( N_2-\frac{N_1^2}{1-N_3} \right)^{1/2} \Big\{ (h_0 \Delta \alpha) + ( \Delta h) (\Delta \alpha)\Big. \nonumber\\
    & & \hspace{1.2cm}\left. - \frac{N_5(1-N_3)-N_1 N_6}{N_2(1-N_3)-N_1^2}\,\frac{h_0}{k}(\Delta \alpha)^2\right.\nonumber \\
    & & \hspace{1.2cm} \Big.+ i (k \Delta h)\, (h_0 \Delta \alpha) \Big\}, 
\end{eqnarray}
where
\begin{eqnarray}
    & & N_4=k^2\, \int\,\dd x |f(x)|^2 x^2 \tanh (\alpha_0 x)\, \text{sech}^2 (\alpha_0 x), \nonumber \\
    & & N_5= k^3\,\int \dd x |f(x)|^2\, x^3 \tanh (\alpha_0 x)\, \sech^4 (\alpha_0 x), \\
    & & N_6=k^2\, \int\,\dd x |f(x)|^2 x^2 \tanh^2 (\alpha_0 x)\, \text{sech}^2 (\alpha_0 x). \nonumber
\end{eqnarray}

\subsection{Comparison for Example II of the classical and quantum Fisher information matrix in the limit $\Delta h$, $\Delta \alpha \rightarrow 0$.}
The elements of the QFIM are
\begin{eqnarray}
& & F_{11}^Q=4 \Omega_{11}=4k^2(1-N_3) \nonumber \\
& & F_{22}^Q=4\Omega_{22}=4N_2 \\
& & F_{12}^Q=F_{21}=4  \Omega_{12}=k N_1.\nonumber
\end{eqnarray}
Using the detection probabilities $p_i$, the elements of the Classical FIM are:
\begin{eqnarray}
& & F_{11}^{C}=\frac{1}{p_0} \Big( \frac{\partial p_0}{\partial h}\Big)^2+\frac{1}{p_1} \Big( \frac{\partial p_1}{\partial h}\Big)^2+\frac{1}{p_2} \Big( \frac{\partial p_2}{\partial h}\Big)^2   \\
& & \rightarrow \frac{1}{p_1} \Big( \frac{\partial p_1}{\partial h}\Big)^2=\frac{\Big[ 2k^2(1-N_3) (\Delta h) \Big]^2}{k^2(1-N_3) (\Delta h)^2} = 4 k^2 (I-N_3) = F_{11}^Q \nonumber\label{FQ11}
\end{eqnarray}
\begin{eqnarray}
& & F_{22}^{C}=\frac{1}{p_0} \Big( \frac{\partial p_0}{\partial \alpha}\Big)^2+\frac{1}{p_1} \Big( \frac{\partial p_1}{\partial \alpha}\Big)^2+\frac{1}{p_2} \Big( \frac{\partial p_2}{\partial \alpha}\Big)^2   \\
& & \rightarrow \frac{1}{p_1} \Big( \frac{\partial p_1}{\partial \alpha}\Big)^2+\frac{1}{p_2} \Big( \frac{\partial p_2}{\partial \alpha}\Big)^2 =\frac{1-N_3}{N_1^2} \frac{4 N_1^4}{(1-N_3)^2}\nonumber\\
& & \hspace{1cm}+4\Big( N_2-\frac{N_1^2}{1-N_3} \Big)    =4N_2=F_{22}^Q \nonumber\label{FQ22} 
\end{eqnarray}
\begin{eqnarray}
& & F_{12}^{C}=\frac{1}{p_0} \Big( \frac{\partial p_0}{\partial h}\Big)\, \Big( \frac{\partial p_0}{\partial \alpha}\Big)+\frac{1}{p_1} \Big( \frac{\partial p_1}{\partial h}\Big)\, \Big( \frac{\partial p_1}{\partial \alpha}\Big)  \nonumber \\
& & \hspace{1cm}+\frac{1}{p_2}\Big( \frac{\partial p_0}{\partial h}\Big)\, \Big( \frac{\partial p_0}{\partial \alpha}\Big)  \rightarrow \frac{1}{p_1} \Big( \frac{\partial p_1}{\partial h}\Big)\, \Big( \frac{\partial p_1}{\partial \alpha}\Big) \nonumber \\
& & \textcolor{black}{\rightarrow}\frac{1}{p_1} \Big[ 2k^2(1-N_3) (\Delta h) +2kN_1 (\Delta \alpha) \Big]\times \\
& & \hspace{2cm}\Big[2kN_1 (\Delta h)+ 2\frac{N_1^2}{1-N_3} (\Delta \alpha)  \Big]  \nonumber \\
& & \hspace{1cm}=\frac{1}{p_1} \Big[ 8k^2 N_1^2 (\Delta h) (\Delta \alpha) +4k^3N_1 (1-N_3) (\Delta h)^2 \nonumber \\
& & \hspace{1.2cm}+ 4k \frac{N_1^3}{1-N_3} (\Delta \alpha)^2 \Big]=\frac{4kN_1}{p_1}\, \Big[ 2k N_1 (\Delta h) (\Delta \alpha)  \nonumber \\
& & \hspace{1.2cm}+k^2 (1-N_3) (\Delta h)^2 +  \frac{N_1^2}{1-N_3} (\Delta \alpha)^2 \Big]\\
& & \hspace{1.2cm}= 4 k N_1 \frac{p_1}{p_1}=4kN_1=F_{12}^Q \label{FQ12} 
\end{eqnarray}

Note that Eqs. (\ref{FQ11})-(\ref{FQ12}) imply that a measuring scheme based on projection onto the modes $\ket{\gamma_k}$ allows to saturate the CR lower bound for estimation precision since $F=F^Q$.

\bibliography{references}

\end{document}